# Exploiting Acceleration Features of LabVIEW platform for Real-Time GNSS Software Receiver Optimization

Erick Schmidt, *The University of Texas at San Antonio*
David Akopian, *The University of Texas at San Antonio*

**BIOGRAPHY**

Erick Schmidt received his B.S. in electrical engineering from Monterrey Institute of Technology and Higher Education, Monterrey, Mexico, in 2011 and his M.S. degree from The University of Texas at San Antonio (UTSA), San Antonio, Texas, United States, in 2015. He is currently a Ph.D. candidate in the Department of Electrical and Computer Engineering, at the University of Texas at San Antonio (UTSA). His research interests include implementation of platforms for software-defined radio algorithms for fast prototyping, WLAN indoor localization systems, and interference mitigation techniques for Global Navigation Satellite System (GNSS). He is a graduate student member of IEEE and ION.

David Akopian is a Professor at the University of Texas at San Antonio (UTSA). Prior to joining UTSA he was a Specialist with Nokia from 1999 to 2003. From 1993 to 1999 he was a staff member at the Tampere University of Technology, Finland, where he received his Ph.D. degree in 1997. Dr. Akopian's current research interests include signal processing algorithms for communication and navigation receivers, and implementation platforms for software-defined radio, and mHealth. Dr. Akopian is a Senior Member of IEEE, member of the Institute of Navigation (ION), and Fellow of National Academy of Inventors.

**ABSTRACT**

This paper presents the new generation of LabVIEW-based GPS receiver testbed that is based on National Instruments' (NI) LabVIEW (LV) platform in conjunction to C/C++ dynamic link libraries (DLL) used inside the platform for performance execution. This GPS receiver has been optimized for real-time operation and has been developed for fast prototyping and easiness on future additions and implementations to the system. The receiver DLLs are divided into three baseband modules: acquisition, tracking, and navigation. The openness of received baseband modules allows for extensive research topics such as signal quality improvement on GPS-denied areas, signal spoofing, and signal interferences.
The hardware used in the system was chosen with an effort to achieve portability and mobility in the SDR receiver. Several acceleration factors that accomplish real-time operation and that are inherent to LabVIEW mechanisms, such as multithreading, parallelization and dedicated loop-structures, are discussed. The proposed SDR also exploits C/C++ optimization techniques for single-instruction multiple-data (SIMD) capable processors in software correlators for real-time operation of GNSS tracking loops. It is demonstrated that LabVIEW-based solutions provide competitive real-time solutions for fast prototyping of receiver algorithms.

**1 INTRODUCTION**

The launch of Global Navigation Satellite Systems (GNSS), and specifically in the US, Global Positioning System (GPS), have enabled various location-based services which have generated widespread studies in related positioning methods, baseband algorithms and techniques, among others [1]. Availability of an accurate source of user position, velocity, and time (PVT) has impacted technologies such as wireless communications, medical and military equipment, ground transportation, commercial devices, etc.
Typically, standard GPS receivers operate nominally in open-sky environments, and are challenged by signal blockages inside buildings, urban canyons, and underground locations. Also, malicious spoofing signals as well as interference may obstruct GPS signals and eventually interrupt location-based services (LBS) which rely on GPS availability. Extensive engineering and research effort is directed towards these challenges to achieve more robust operation of the receivers by increasing their coverage to as many denied areas as possible. GPS signals themselves are hidden below the noise floor; therefore, improvements increase receiver sensitivities by providing access to terrestrial signaling channels. Using ground communication links, receivers can retrieve orbital data of satellites, and receiver coarse position and time estimates from wireless networks which significantly enhances GPS receivers' robustness. This approach is called Assisted GPS (A-GPS) [1]-[3] and is estimated to improve start-up time to obtain a PVT solution, as well as a sensitivity gain as much as 20 dB when used in combination with other advanced techniques such as parallel correlators. The A-GPS solution is standardized for telecommunication and cellular networks in terms of defining logistics of providing relevant assistance data to the user, such as coarse PVT information. A-GPS is also a recommendation by FCC E911 mandate as support for emergency services. Other advances in terms of receiver coverage increase are made in terms of better immunity of GPS receivers against interferences and malign spoofing signals [4]-[7]. Despite this progress, GPS operation is still denied in many indoor areas and other weak signal environments. Also, even with many reported spoofing mitigation methods [4], these interference and spoofing techniques continuously evolve and therefore bring new challenges for GPS receivers.

Researchers succeed to improve the performance and coverage of GPS receivers by addressing continuously evolving spoofing threats. For these advances, researchers required proper instrumentation and software to support their efforts. As such, software-defined radio (SDR) GPS receiver solutions become popular because of providing full control and openness of receiver internal operations and algorithms. Therefore, researchers can use these solutions to integrate and test new algorithms and methods for improved performance and general robustness, as well as mitigation techniques. SDRs are composed of peripheral RF front-ends, which receive sampled data, and then apply signal processing using general-purpose computing resources such as CPU processors, as well as integrated accelerator and reprogrammable devices such as DSPs and FPGAs.

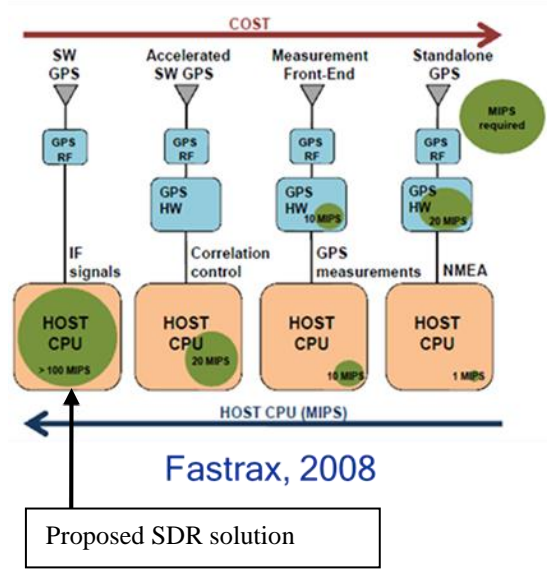

Figure 1. Implementation cost of a GPS receiver.

The basic understanding for software-defined radio is implementation of conventional communication baseband modules in software; and a hardware part that takes care of receiving real-time signals and converting them into the digital domain for post-processing. Commercial receivers typically provide only application programming interface (API) access, which limits internal algorithm manipulations. They are also very costly (could be more than $20K), for a complete solution (front-end and software module). Figure 1 shows a typical scenario on how SDR systems have evolved over time when compared against their cost and hardware-to-software transfer in the receiver. One can see that standalone GPS receivers are the costliest and since most of the implementation is done in hardware, they are considered a closed system. Therefore, one cannot access or modify their algorithms for testing and fast prototyping, as opposed to the almost-all software SDR scenario. SDR gives us: reconfigurability (can configure for different frequencies, sampling rates, etc.), flexibility (adapt functionality in real time, trade-off speed, power, by using different settings), reusability (re-use certain modules for other tasks, future multi-constellation GNSS development potential, etc.), portability (front-ends are compact nowadays as well as the host platform used), interoperability (software is typically deployed in several host PC form factors as well as embedded hardware platforms), as opposed to hardware (typically purpose-specific and standalone, no access to RF chain modules for experimentation). This

paper proposes an SDR with hardware which was selected so that cost is effective for its implementation as well as being able to access conventional GPS baseband modules for research and testing.

GPS SDR solutions can be currently found in various configurations, but have certain limitations such as a lack of real-time operation, cost-effectiveness, compatibility, openness of algorithms, among others. These limitations occur in such combinations such that for instance, a real-time receiver could have closed/non-modifiable baseband modules for research purposes, or could be an expensive proprietary solution which offers only API access to functions. Another limitation could be an open-source solution which exploits real-time operation, but is highly dependent (or not compatible in other words) to certain software frameworks for it to function properly. Such SDR options are the following: (a) proprietary solutions providing API-only access for 3$^{rd}$ party development [8], which are typically compiled and fully optimized in C/C++ language for generic use; (b) open-source snapshot C/C++ solutions such as [9], therefore lacking real-time operation; (c) open-source real-time C/C++ solutions that depend on frameworks such as GNU Radio [10],[11]; (d) MATLAB/Simulink-based solutions mainly for theoretical research, typically not real-time and even slower than C/C++ versions [12]; (e) academic solutions employing hybrid development environments, such as C/C++ solutions integrated in LabVIEW [13]-[16], or even with leveraging FPGA and DSP accelerator hardware [17].

This paper presents the new generation of LabVIEW-based real-time GPS receivers. It also provides in-depth study of GPS SDRs acceleration using inherent LabVIEW mechanisms, such as multithreading, parallelization and dedicated loop-structures. It also exploits C/C++ optimization techniques for single instruction multiple data (SIMD) capable processors in software correlators for real-time operation of GNSS tracking loops. It is demonstrated that the proposed LabVIEW-based solution provides a competitive real-time receiver for fast prototyping of research and optimization algorithms.

**2 HARDWARE TESTBED**

The development and testbed platform is implemented by Software Communications and Navigation Systems (SCNS) Laboratory at the University of Texas at San Antonio (UTSA). Figure 2 shows the full hardware development testbed that was used in this paper, which includes an SDR GPS system which acts as the receiver and a GPS simulator which acts as a generator/transmitter of GPS signals. The former is the novelty described in this document and the main development effort, and the latter is a product from National Instruments that is used for specific laboratory testing and research work. The A-GPS portion of the testbed seen in Figure 2 is not discussed in this paper, but its implementation can be found here [13]. This section will describe the full development hardware testbed components used in this SDR system and their specifications.

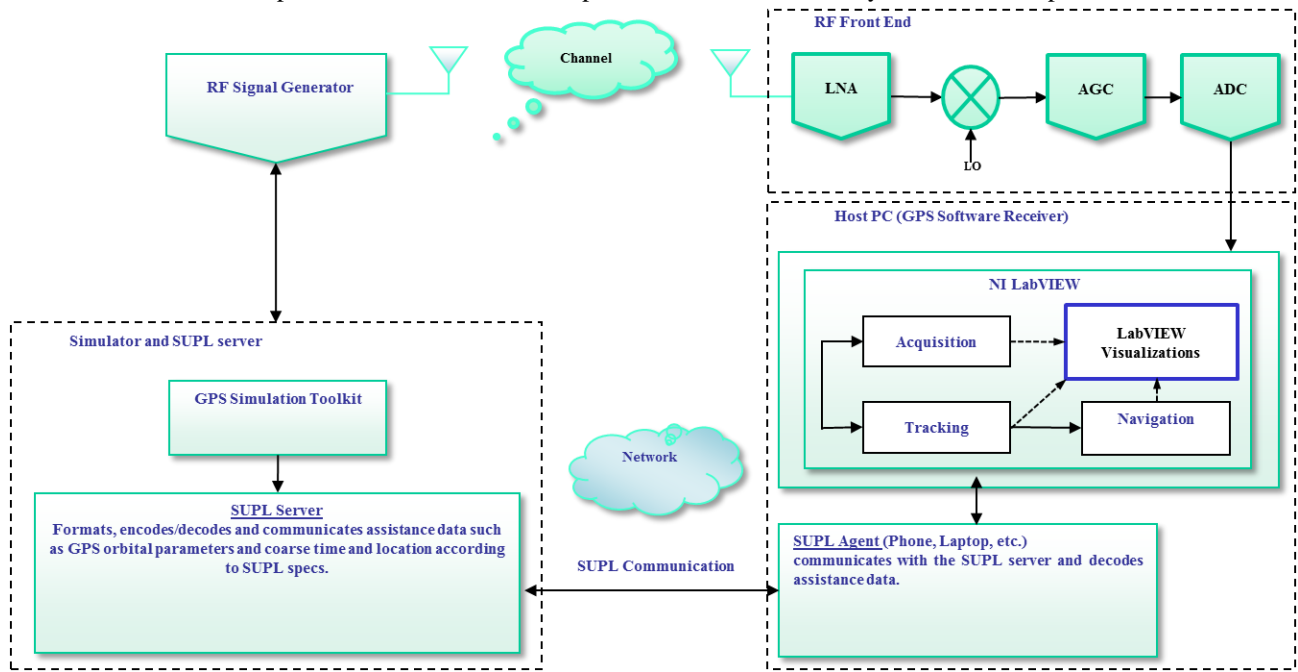

Figure 2. UTSA's development and testing platform for GPS SDR including RF Front-End, GPS simulator, SDR receiver, A-GPS support, host PC, and software platform.

**2.1 GPS simulator (transmitter) hardware**

National Instruments provides an add-on toolkit in LabVIEW environment for generating GPS signals. This system is the GPS Simulator Toolkit v2.0 [15]. Figure 3 shows the hardware configuration and an actual picture of the setup used.

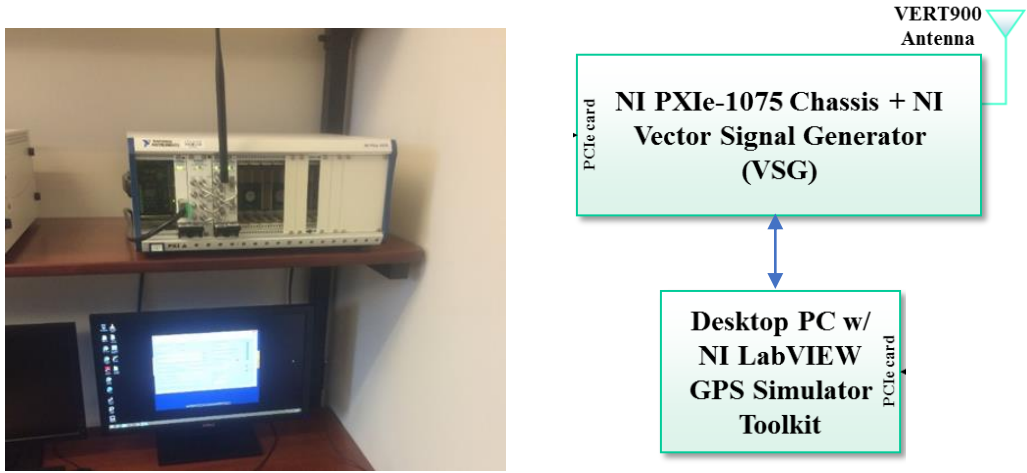

Figure 3. NI GPS Simulator Toolkit and NI Chassis signal generator hardware.

This system is installed on a host PC in the SCNS laboratory where several tests are performed. The transmitter's RF front-end consists NI PXIe-5673 RF vector signal generator (VSG) paired with a VERT400 antenna. The GPS signal is output from the software simulation toolkit which is highly configurable. Using this toolkit, the VSG generates GPS signal which is upconverted to L1 frequency and transmitted via the antenna. It can simulate various channel distortions as well as enable concurrent generation of up to 12 satellite signals (PRNs). Signal powers can also be adjusted independently for each satellite. The simulator also supports signals of a wide area augmentation system, trajectory scripts, on-the-fly parameters, and stored files for recording and retransmitting test signals. The simulator uses orbital data, i.e., ephemeris and almanac data file inputs. Therefore, one can generate static as well as dynamic tests for the GPS SDR to decode, based on configuration inputs mentioned previously, even though the user might not actually be moving. This simulator was used extensively for general and specific tuning of the SDR system as well as for algorithm testing and hardware/software decision making in the beginning, performance improvements, speed improvements, and other useful modifications.

**2.2 GPS receiver hardware**

Although most of the receiver algorithms are implemented in software mode, some portion of the SDR system requires a hardware part, which oversees collecting raw analog signals from the air and converting them into digital domain samples (since software modules only understand digital data) to be fed into the software system. The hardware for this SDR testbed was selected to maintain portability. This in term is used for outdoor static and dynamic testing scenarios as well as possible drive tests. Figure 4 shows the main components of the hardware setup.

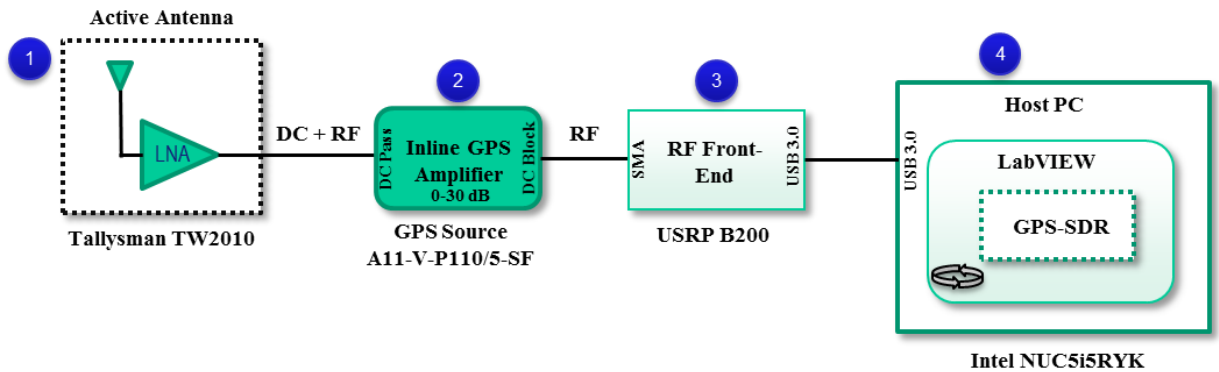

Figure 4. Hardware setup for the SDR system.

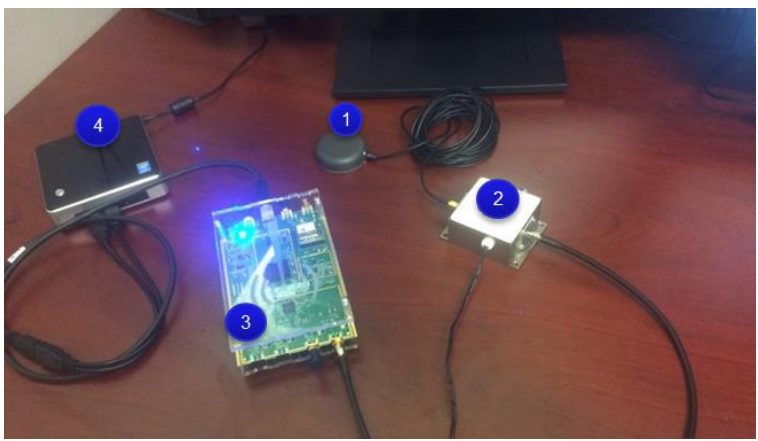

Figure 5. Actual image of the SDR components (numbered).

There are in total four components that comprehend the hardware part of the receiver: An active antenna, an in-line amplifier, an RF front-end, and a multi-purpose processor host PC where the software part of the SDR is found. The first three components are in charge of obtaining raw GPS signals over the air, which are in turn sent to the host PC where they are processed in the form of digital samples. The choice of these components was made always keeping in mind portability and mobility, as well as reusability for future additions to the system such as multi-constellation capabilities. Figure 5 shows an image of these four components which give an idea of the size and weight (see Figure 4 as well to identify each numbered component). For the front-end and the host PC, tables 1 and 2 summarize the capabilities of these hardware parts, which are well enough for GPS L1 coverage. The front-end of choice is the Ettus USRP B200.

**Table 1. Host PC specifications**

| Host PC | Intel NUC5i5RYK |
|---|---|
| CPU | Intel Core i5-5250U @ 1.6-2.7 GHz, dual-core, 3 MB cache, 15 W |
| RAM | 16 GB RAM DDR3L @ 1600 MHz |
| Storage | 240 GB M.2 SSD |
| Operating System | Windows 7 Ultimate (64-bit) |
| Dimensions | 115 mm × 111 mm × 32 mm |
| Weight | ~400 g |

**Table 2. RF front-end specifications**

| RF Front-End | Ettus USRP B200 |
|---|---|
| RF Coverage | 70 MHz to 6 GHz |
| Bandwidth | 200 KHz to 56 MHz |
| ADC Resolution | 12-bit |
| Oscillator | GPSDO (OCXO), frequency stability: ±25 ppb |
| Interface | SuperSpeed USB 3.0 |
| Dimensions | 97 mm × 155 mm × 15 mm (Board only) |
| Weight | 350 g (Board only) |

The front-end has the following RF components: a multi-stage low noise amplifier (LNA), a local oscillator (LO), an automatic adaptive gain controller (AGC), a low-pass filter (LPF), and a state-of-the-art analog-to-digital converter (ADC) [18]. The ADC is considered state-of-the-art since it performs direct conversion to baseband, as opposed to the classical heterodyne receiver which is multi-stage, and thus requiring more RF components such as filters for the intermediate frequency (IF). The ADC resolution outputs 12-bit samples, which are then scaled up to 16-bit by the DSP inside this device for a more generic 2-byte sample size that can be understood by a host PC. The ADC also has internal configurable decimation filters to lower this 16-bit native output to an 8-bit sample format, which improves throughput between the interface and the host PC via USB 3.0. The ADC sampling rate can achieve speeds up to 56 Msps. Since the GPS SDR system was tested for 5 MHz sampling rate in IQ interleaved format, the total throughput through the USB cable can reach up to 10 MB/s for INT8 data format. An oven-controlled crystal oscillator (OCXO) GPS-disciplined clock (GPSDO) [19], commonly known as a GPS clock was used for frequency accuracy (see Table 2). This component can be installed separately on the board (it mounts directly by attaching pins) as opposed to the built-in temperature controlled crystal oscillator (TCXO) which has lower frequency accuracy. GPS signals require high-accuracy clocks to avoid phase uncertainties for satellite navigation applications on tracking and positioning [20].

**3 SOFTWARE ARCHITECTURE**

This section describes software functionalities of the SDR system. GPS relies on three conventional baseband processing modules: acquisition, tracking, and navigation. These modules were implemented in C/C++ language by creating main-purpose reusable functions that were compiled with full optimization on 64-bit version for speed, performance, and high memory allocation if need be. These functions were compiled as dynamic link libraries (DLL). These DLL libraries are afterwards accessed via the software platform LabVIEW that takes care of the interface and data flow of the SDR system. Since baseband modules used are based off conventional GPS demodulation processes, two reference receivers were used for the development: *fast-gps* [9], and *softGNSS* [12]. These reference receivers are typically not real-time; therefore, they were modified algorithmically, split into independent functional modules, and finally used to real-time mode based on DLL functions accessed via a controlled LabVIEW data flow environment. The SDR is considered interoperable since it can operate on any host PC that has a 64-bit Windows Operating System (has been tested in Windows 7 and Windows 10 successfully), LabVIEW 2016 64-bit version, and the latest NI-USRP driver.

The receiver is capable of tracking up to 12 channels in real-time mode. This number of channels is hard-coded on the receiver internals and it's based on real scenarios where no more than twelve GPS L1 satellites are visible on the sky at any given moment. The system was compiled and developed in a 64-bit environment. There is a requirement for a 64-bit processor, operating system, as well as LabVIEW 64-bit version for this SDR to work. The system can configure and use any sampling rate as long as the front-end is compatible, although in this SDR system the sampling rate used was 5 MHz as it showed a good quality and allowed extra overhead computational space for further testing and research.

Figure 6 shows the main front-panel of the software receiver. There are two main functionalities for this receiver: real-time operation and simulation (offline) mode. For the real-time operation, the front-end needs to be compatible and properly detected by the NI-USRP driver at the time of execution. Also, the front-end needs to have an active antenna as well as an amplifier to drive this antenna. An extra requirement would be an extra active antenna to use the GPSDO clock in locked mode for enhanced precision. For the offline mode, a binary file must be used. The advantage of offline mode, as opposed to real-time operation, is that there a variety of recording files available, with different sampling rates, sampling type (I vs I-Q samples), and a possibility to use intermediate frequency (IF) recordings such as those from the GN3S front-end [23]. In offline mode, extensive debugging can be used since the software is not dependent on real-time operation.

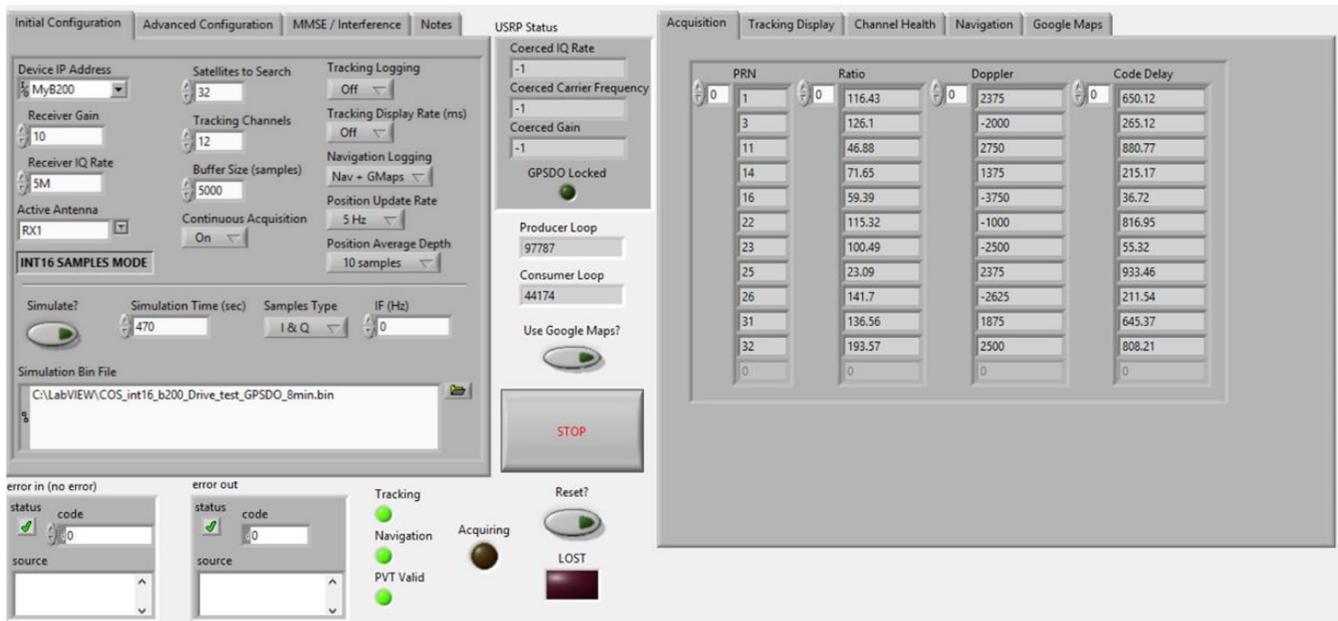

Figure 6. Front panel of the working SDR receiver in LabVIEW. *Left* shows configuration parameters, and *right* shows acquisition visualization outputs.

### 3.1 GPS L1 baseband processing

Conventional GPS processing is widely covered in literature [1], [12], [21], [22]. The GPS system consists of 32 satellites orbiting the Earth, which transmit a signal on the L1, L2, and L5 carrier frequency bands. The payload data contained in each satellite vehicle (SV) transmission is modulated into three layers: a BPSK modulation, a spreading sequence, and the carrier wave. The transmitted GPS signal from each satellite for the L1 band can be modeled as seen in Equation (1):

$$s_{L1}(k) = \sqrt{2P_C}\,d(t)\,c(t)\cos(2\pi f_{L1}t + \theta_{L1}) + \sqrt{2P_Y}\,d(t)\,y(t)\sin(2\pi f_{L1}t + \theta_{L1}) \qquad (1)$$

There are two parts of the signals transmitted, each one being on the in-phase and quadrature channels respectively. The in-phase (cosine) branch corresponds to the C/A signal which stands for coarse acquisition and is also called the civilian signal. This signal is freely available for the public. The other branch (quadrature) transmits the P/Y signal which is for military use and is restricted for civilian use. $P_C$ and $P_Y$ represent the signal power for these two signals, d(t) is the BPSK modulated navigation data, c(t) and y(t) represent the ranging codes for the C/A and P/Y signals respectively, and the carrier wave is represented by the sine and cosine, with a carrier frequency centered on $f_{L1}$ as well as an initial phase $\theta_{L1}$. This paper focuses on the L1 C/A signal.

GPS relies on direct sequence spread spectrum (DSSS) signaling. This means the navigational data is combined with a higher data rate signal bit sequence for spreading the spectrum. These bits are called chips. The ranging code c(t) for the C/A L1 signal is used to spread the signal in bandwidth, thus creating a noise-resistant signal. This sequence is often called a pseudo-random noise (PRN) code, and consists of a known combination of 1023 chips. The C/A code is transmitted at a rate of 1.023 Mcps (chips per second) and consists of a sequence of +1's and -1's. This means that one period of the 1023 chip sequence has a duration of 1 millisecond. The navigation data is BPSK modulated and transmitted at a rate of 50 bps, which is combined with the PRN code, and finally the carrier wave. Each navigation data has a length of 20 ms, therefore there are 20 PRN code periods in each bit. Each satellite (1-32) has a unique PRN code. The PRN codes have special properties of orthogonality, thus are barely interfering among them. The navigational data which is transmitted by each SV combined with its respective PRN code, contains required information for the position solution.

The final goal of a GPS receiver is to provide user position, velocity, and time (PVT) information. To obtain the PVT, receivers require two parameters: range measurement and navigation data. Range measurements are extracted by synchronizing a locally generated replica of the PRN code with the received signal. This synchronization is performed in time, by aligning the signal and replica to have the same so-called code-phase, and in frequency, by compensating for residual Doppler modulation after carrier wipe-off. PRN signals are aligned when the edges of the code periods are aligned. The synchronization is typically performed in two phases: acquisition (coarse) and tracking (fine). After time and frequency synchronization the navigation data are simply obtained through PRN code wipe-off and integration. Figure 7 shows a basic demodulation scheme.

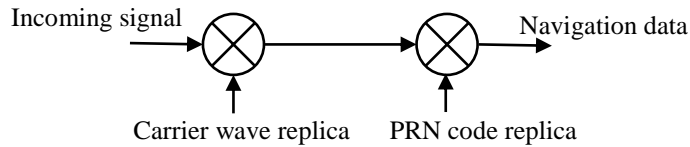

Figure 7. A basic demodulation scheme.

The navigation data contains orbital and clock parameters which are used in the computation of user position. In addition, navigation data includes a time-stamp message which can be used to compute range measurements by a time-difference of arrival (TDOA). The identification of the time-stamps in the received signal along with the code-phase measurements is used to obtain time of the signal transmission and then user-to-satellite ranges. After obtaining ranges from several satellites, the user is able to use a trilateration technique for a position calculation. Typically, least-mean square (LMS) solution is used to compute user position and requires at least 4 satellites.

**3.2 Software functionality**

The main functionality of the receiver is based on collecting raw digitized samples from the front-end and post-processing them in real-time for an end result which is a position solution. LabVIEW is based on programs called virtual instruments (VIs) which can be programmed independently and later on attached to work either as main or sub VIs. There is a main VI where all the upper layer functionality is held such as data flow and main execution of loops. Figure 8 shows the main components of the receiver in LabVIEW environment and how the data flow interacts with the receiver and the front-end.

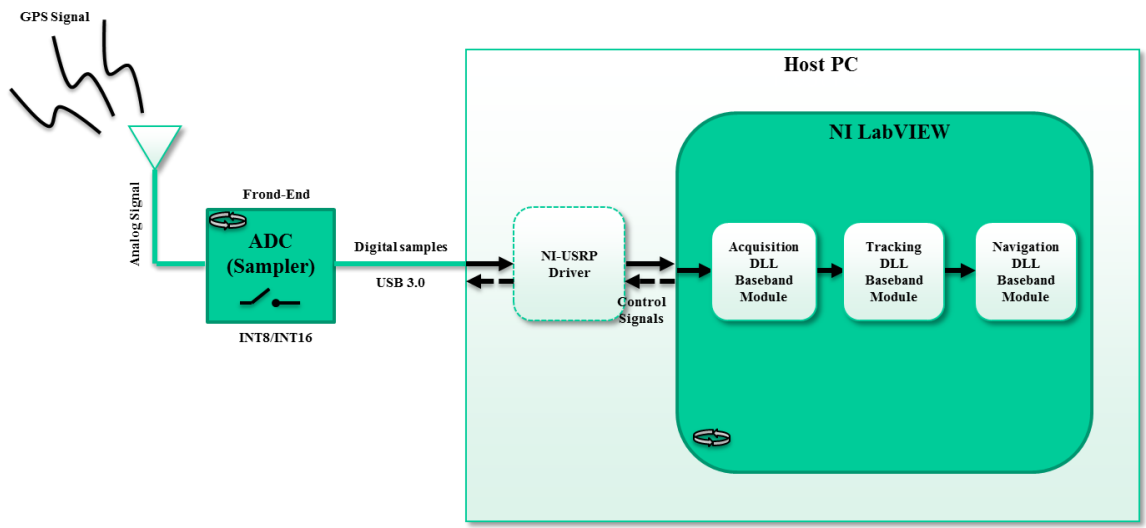

Figure 8. Main SDR functionality.

The GPS signal is converted to samples in the digital domain and fed via USB 3.0 interface to the host PC. There, the NI-USRP driver collects the samples in blocks of data and sends them to the main LabVIEW VI, which eventually passes these samples to the DLL baseband modules. Figure 9 shows a more detailed view of the internal flow of data inside the main VI.

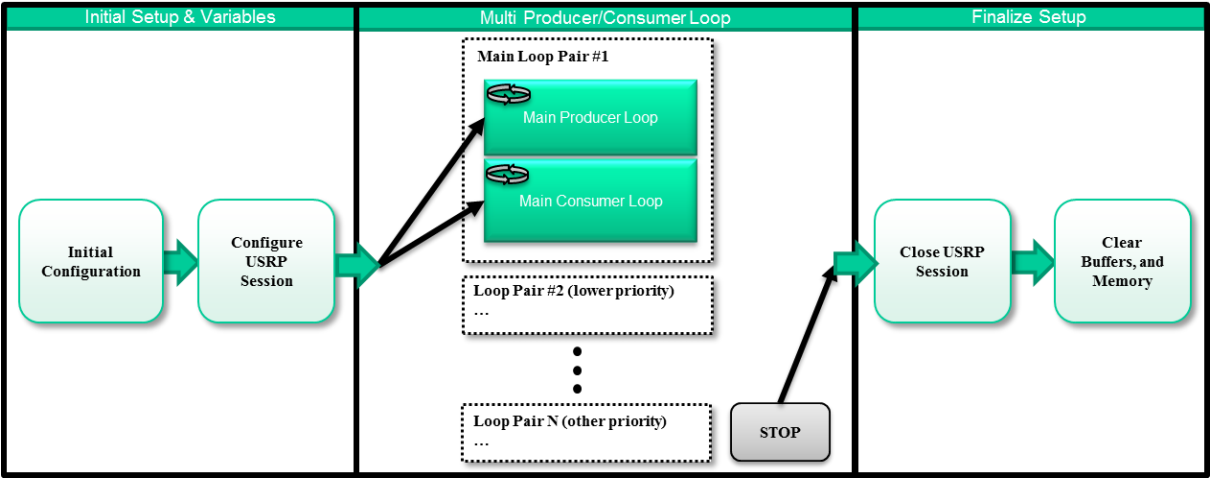

Figure 9. Internal data flow on LabVIEW SDR.

The flow inside the main VI consists of three stages: Initialization of variables, the main loop (Multi-producer/consumer loop), and the finalize setup and close session. The initialization part is in charge of initializing variable flows for system settings which contain configuration variables for the current GPS session, and a twelve channel parameters array. Each channel contains a cluster of variables containing crucial tracking loop data, channel health data, and other essential information. After these two main variable structures are initialized, USRP session is configured and initialized. This includes setting up a session with to the front-end to setup parameters such as sampling rate, block size, data format, reference clock, etc.

The main processing of the receiver works inside the multi-producer/consumer loop section. The producer/consumer loop is a known design structure used for current processing of independent tasks. Each loop is a "while" loop that run until receiver issues a stop command. The producer loop can be seen as a buffer that inputs data into a memory queue. Then the consumer loop dequeues and processes this data. This is achieved by LabVIEW's Queue/Dequeue built-in function blocks. Although there are multiple of these loop pairs, the main producer and consumer loops are the most important and it is where all the baseband processing occurs. The consumer loop also acts as a producer loop for several other lower priority tasks. It enqueues data for mainly visualization loop outputs which are lower priority. These extra secondary loop pairs can be added for different priority tasks in the receiver, but are mainly for visualizations so their priority can be changed on-the-fly and therefore offload computational overhead for the main loop.

**3.3 Main Producer/consumer loop**

Figure 10 shows the main producer loop. This consists of while loop that runs until the receiver is stopped. It has a single subVI function inside it which is in charge of requesting data from the front-end.

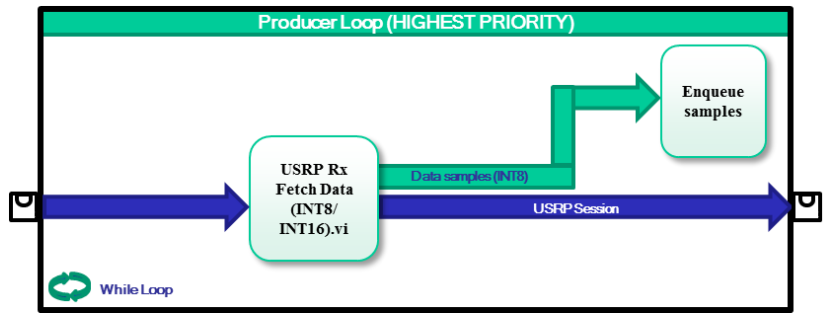

Figure 10. Main producer loop detailed.

One can see a pipe which contains the *USRP Session* connection; a reserved data wire containing the current configuration and session of the USRP for the receiver. This wire is created when the USRP is initially configured in previous section. The Enqueue samples oversees that samples fetched from the USRP Rx Fetch Data (INT8) subVI are sent to the queue buffer. This Enqueue function takes the data samples and sends it to main memory in the form of a lossless queue. This data enqueue occurs in an inherent multi-threading environment based on LabVIEW's diagram flow. The *USRP Session* wire is attached to a shift register (small black box on both ends of the loop) achieving storage of last values and recovering them for next iteration.

Figure 11 shows detailed consumer loop and how the data flow is handled. Concurrently and in a parallel graphical flow (based on LabVIEW multi-threading and parallelism feature), the consumer loop collects data from the queue (Dequeue) generated previously for the main producer. The main consumer loop contains the main C/C++ modules and therefore all the baseband blocks compiled as DLLs. Main data samples post-processing occurs in every iteration of this loop, which processes the data from left to right as seen in Figure 11. The first step executed in this loop is the Dequeue operation which checks for a block of data samples. It takes this block of samples and sends it to the ACQ baseband module as well as the Tracking module for post-processing.

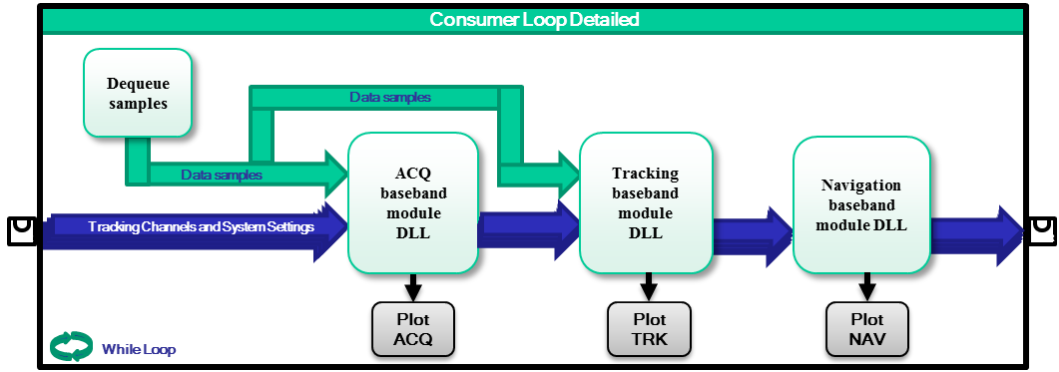

Figure 11. Main consumer loop detailed.

The ACQ module contains the acquisition and therefore, if the receiver is in the ACQ state, it will try to fill its buffer (*acq_buf*) before performing the acquisition search algorithm. This buffer is pre-initialized based on the configuration of the integration length (ms). The ACQ baseband module can perform an initial acquisition when there are currently no channels being tracked; this is known as "cold start". The ACQ module can also perform a continuous acquisition which occurs in parallel with tracking by trying to find new channels for improving the number of satellites for a better navigation solution. Since the acquisition execution is computationally heavy, only when there are a small number of channels tracked i.e. less than four, another acquisition is triggered. If the acquisition finds satellites to track, the receiver changes its state to tracking. Tracking stage begins tracking the coarsely found channels by the acquisition algorithm continuously. Once in tracking, and as long as continuous acquisition is not required, data samples sent to the ACQ module are discarded and the module itself is skipped on every iteration. Tracking module continuously (fine) synchronizes to incoming signal and attempts to collect navigation data. Table 3 lists tracking loops configuration used in this module, and is considered a conventional tracking mode.

**Table 3. Tracking loop configuration parameters for baseband module.**

| Parameter | GPS L1 |
|---|---|
| Code loop type | $2^{nd}$ order DLL |
| Early-late spacing | ½ chip |
| Code loop discriminator type | Normalized Early minus Late Envelope |
| Code loop aid | Carrier aiding |
| Carrier loop type | $2^{nd}$ order PLL + $1^{st}$ order FLL |
| Carrier loop discriminator type | PLL: *atan*; FLL: *atan2* |

Once the Ephemeris data has been decoded from tracking, the state machine jumps into navigation. Navigation stage is in charge of estimating LMS positioning based on the tracking data, pseudoranges, ephemeris data, and other parameters. A minimum of 4 channels is needed for a position solution. Acquisition and Tracking modules extract relevant information and send it to visualizations in real-time. The Navigation module also extracts positioning information for the visualization tabs, and sends it to visualizations.

**3.4 LabVIEW DLL integration**

Dynamic Link Libraries contain optimized C/C++ functions previously compiled for each GPS baseband module. Independent modules with built-in function calls in their respective libraries. Full optimization is done at compilation by Visual Studio 64-bit compiler. Data flow is handled by LabVIEW and it can call these functions based on the *Call Library Function Node* VI. Figure 12 shows this VI and how to configure it.

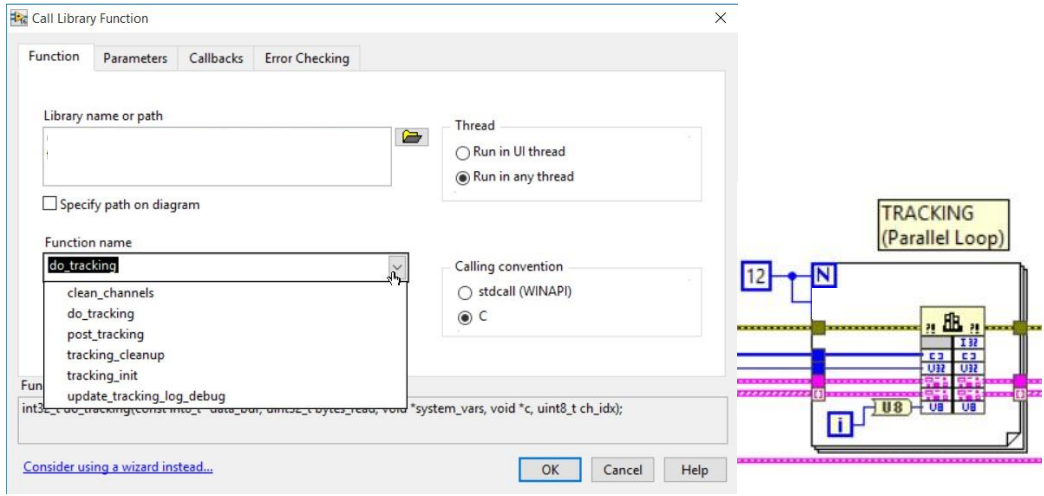

Figure 12. (left) *Call Library Function Node* configuration, and (right) actual VI icon in block diagram.

The Call Library Function Node should be configured to point to a .dll file which contains all the library functions. Once the file is selected, all these functions inside the DLL appear in the function name dropbox. These functions should be specially coded in the Visual Studio solution by using the appropriate calling convention for DLL function calls, as the C/C++ solution should be compiled as a DLL. Figure 12 shows an example of the tracking execution *do_tracking*. Once the function is selected, one should configure the input and output parameters which in this case, for most of the functions it involves the raw data block (bold blue wire), its data size (thin blue wire), the system parameters (thin pink wire), and the array of tracking channel clusters (bold pink wire). The functions inside the DLL use a commonly known way to pass the LabVIEW clusters in a method called: *function call by reference*. This means that the data is not passed directly which would mean to have to make a copy of the data every time these functions are executed; instead, their address in memory is passed to the function so it can access the data directly. This not only avoids duplicating data, it also highly improves execution time in functions. Recent versions of LabVIEW have a very useful compatibility with data types in C/C++, therefore this passing data by reference is easily achieved and integrated onto the SDR system.

**3.5 LabVIEW parallelizable loops**

LabVIEW has a functionality called *parallelizable loops*, which essentially turns a sequential for loop structure into a parallel execution. For this execution to work in parallel, the iterations inside the loop should have no dependency variables or common shared resources. This feature greatly leverages the tracking loops since it treats each channel independently in a common parallel task. Figure 13 shows an example of the implementation of this feature on the tracking loops.

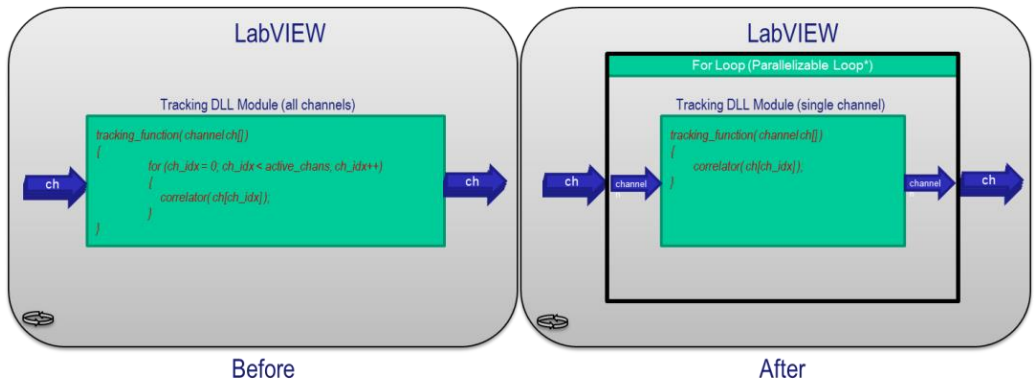

Figure 13. How the parallelizable loop was implemented in LabVIEW.

Previously, the DLL tracking function call took as input an array of all the current tracking channels, either active or not, and inside the C/C++ function a *for-loop* handled all the tracking channels sequentially. Afterwards, the DLL tracking function call was modified so that it processed only one tracking channel at a time. This left way for LabVIEW to use the parallelizable loop feature. One can see in Figure 13 that the for loop is handed to LabVIEW, and now this loop structure can be manually configured to use the parallelizable loop feature, making sure tracking channel variables are independent on one another. Some dependencies are now handled on a post-tracking function, such as receiver clock adjustment based on tracking channel information. This post-tracking is much less computationally intense when compared to the tracking correlators.

**3.6 LabVIEW visualizations**

LabVIEW offers several visualization options that are built-in and can be configured to show specific charts and/or graphs in real-time display. This section will describe the visualizations available in the SDR system, as well as their real-time configuration options. These tabs are found on the right side of the main front panel, and can be visualized in real-time. The display tabs available are: *Acquisition, Tracking Display, Channel Health, Navigation* and *Google Maps*. Some of these visualization tabs offer real-time configuration options for their visualization.

A successful A*cquisition* with 11 SVs found can be seen in Figure 6 on a previous section. One can see which PRN codes were found in the GPS signal (out of 32 known PRN codes), their acquisition ratio, the coarse Doppler frequency found for that satellite signal, and the code delay found which is typically a value between 1 and 1023 chips. These parameters are used for the next stage in the GPS receiver which is tracking the signal for those channels. The *Channel Health* tab is an important visualization because it gives you the status of each tracking channel in real-time, and can give an overall assessment of the stability of the signals and robustness of the algorithms as well. It displays all 12 tracking channels and 8 columns for each channel with crucial information for monitoring them: *PRN, C/$N_0$ (dB-Hz), Carr Lock Ratio, Lock Fail Count, Nav State, ValidPVT* led, and other LEDs not discussed in this paper. Figure 14 shows this visualization tab.

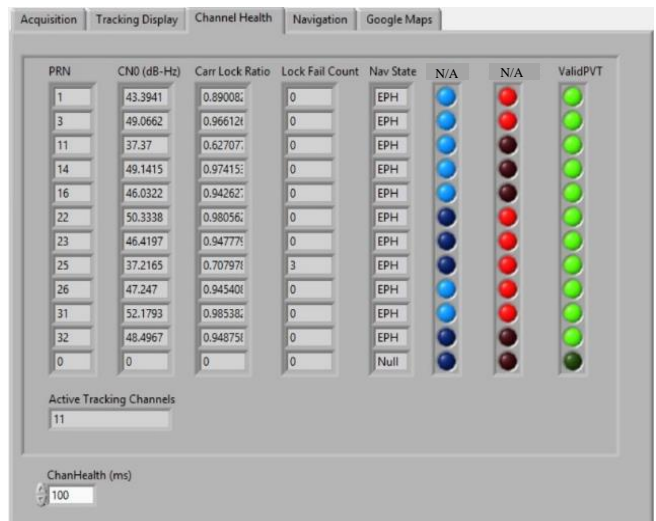

Figure 14. Channel Health tab visualization tab on LabVIEW-based receiver.

For the *Channel Health* tab, the *PRN* represents the satellite vehicle (SV) number that is assigned to that channel (can be 1-32). The *C/N$_0$ (dB-Hz)* stands for carrier-to-noise ratio, and it's defined as the ratio of the carrier power and the noise power per unit bandwidth (dB-Hz) [24]. This measurement is computed inside the tracking loops, based on the signal-to-noise (SNR) ratio and other measurements of the unmodulated carrier wave and noise. The C/N$_0$ calculation is selected from [25] which provides a good approximation with fewer computations. In general, the C/N$_0$ measurement indicates the signal power on that tracking channel as seen by the front-end. The *Carr Lock Ratio* indicator is an estimator to know the carrier lock status based on the estimate of the cosine of twice the carrier phase error [11]. This last estimate has values ranging from 0 to 1, and can be seen as a percentage of how much of the signal power is found in the in-phase arm, i.e. how well is the phase lock loop doing its job by keeping the signal in the I-channel. The *Nav State* shows the status of the navigation data that has been demodulated and processed in each tracking channel. It consists of three states: Null, TOW, EPH. As mentioned in previous sections regarding the GPS data structure, the channel first finds the GPS timestamp (TOW state) value, the receiver collects and processes the Ephemeris data (EPH state) which contains clock and orbital parameters. Only when four satellites or more are in the EPH state, and the *PVTValid* indicator is on, can there be a position solution.

The *Tracking Display* tab shows and updates results in real-time of a selected tracking channel in three different visualization graphs found in the tab. This tab is useful for debugging in real-time how a channel is behaving with respect to the following three outputs: The In-phase Prompt (I Prompt) channel magnitude (top graph), a I vs Q constellation graph (lower left graph), and a Doppler frequency graph output from the PLL (lower right graph). This tab also contains configuration parameters at the bottom that can be modified during execution (on-the-fly). Figure 15 (left) shows this tab. Finally, there are two useful tabs for visualizing the receiver's navigation solution: The *Navigation* tab, and the *Google Maps* tab. The *Navigation* tab displays a graph which plots points from the receiver's navigation solution. These points show the receiver position in Latitude and Longitude (degrees), which is the output of the navigation module, and these points are populated in real-time but have no reference map to pinpoint actual location. The *Google Maps* tab offers the services of a Google Maps API which can be appended to LabVIEW visualizations. The visualization contains configurations such as zoom level, and map type. For this visualization, the host PC must have access to an Internet connection. GDOP and RMS Error can also be seen in these visualizations, which are updated in real-time. A google maps visualization can be seen in Figure 15 (right).

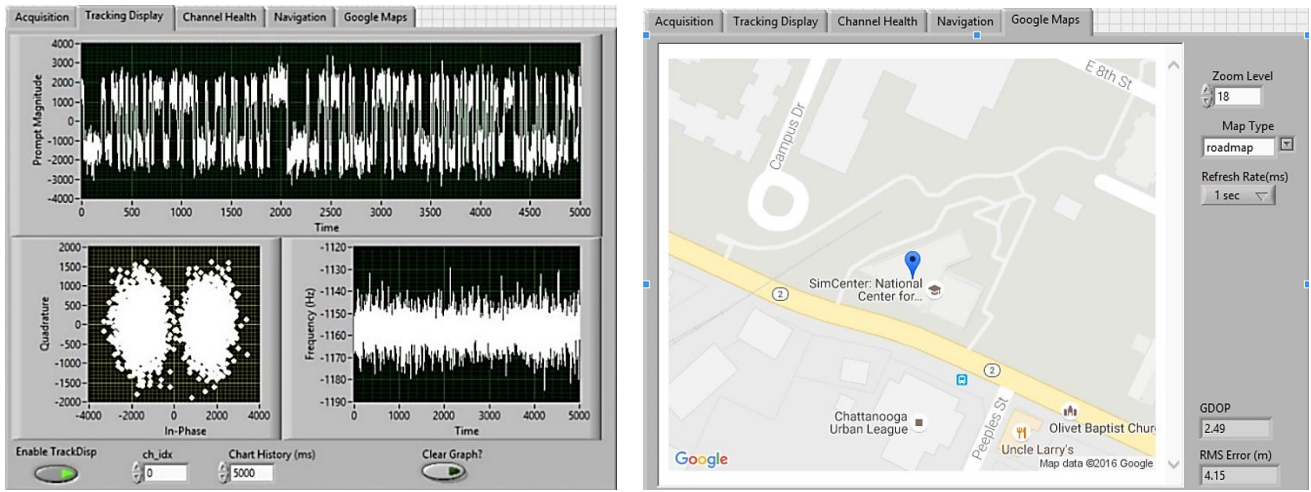

Figure 15. (*Left*) *Tracking Display* tab visualization, and (*right*) *Google Maps* tab visualization on LabVIEW-based receiver.

**3.7 Acceleration factors**

Table 4 summarizes a series of acceleration factors used in the proposed receiver and how they relate to specific technologies. For *algorithmic accelerators*, an advanced acquisition algorithm seen on [26] was implemented as a joint-search FFT space acquisition and compared against conventional PCS acquisition algorithm. Also, for software correlators, a single-input multiple-data (SIMD) was applied, as reported on [27], which provided outstanding gains. These performance gains will be discussed in the next section. Optimized libraries such as FFTW [28] were used for FFT-based algorithms in acquisition and other functions. All LabVIEW features were mentioned in previous sections and the following section will provide acceleration gains. Final acceleration factor with FPGA and DSPs was not implemented in the proposed receiver but it was reported on [16] as previous work.

**Table 4. Proposed SDR acceleration factors**

| Acceleration factor | Source |
|---|---|
| Algorithmic accelerator | Software-based |
| Single instruction, multiple data (SIMD) | C/C++ and Intel processors |
| Optimized libraries | C/C++ libraries |
| Parallel/multi-core scheduling | LabVIEW-based |
| DLL integration | LabVIEW-based |
| Inherited multithreading | LabVIEW-based |
| Data acquisition, data flow control | LabVIEW-based |
| Real-time advanced visualizations | LabVIEW-based |
| FPGA and DSP | Hardware-based |

With regards to parallel/multi-core scheduling, LabVIEW schedules automatically processes to cores based on the host PC capabilities. Therefore, one can use parallel architectures such as multicore CPUs, FPGAs, and GPUs to leverage this feature without the need to write extensive code for parallelism. LabVIEW program structures show, on average, a 25 to 35 percent improvement in execution time because of the inherent parallelism in the block diagram code [29]. Furthermore, LabVIEW offers inherent multithreading optimization which means it automatically adjusts to multicore processor capabilities and memory expansion (it schedules processor threads to each core dynamically). Data acquisition, flow and control is seamlessly handled when using the producer/consumer loop architecture. Multiple producer/consumer loops add control to each thread as one can control data flow speed and refresh rate. Finally, most visualizations are built-in as a graphical user interface (GUI) and can be useful for fast prototyping.

## 4 RESULTS

This section discusses tracking optimization results for the proposed receiver for offline (simulation with a recording file) and online mode (with front-end) of operation. Three most significant acceleration features were considered for comparison purposes of the receiver: multiple-producer/consumer loop feature, parallelizable loop feature, and SIMD feature. Each next version contains previous features, i.e. the SIMD feature contains all previous features. For offline testing, a 300 second recording file which had 12 visible SVs at all time was used for testing. The receiver was tested in simulation mode with this 300 second file and timing was calculated. This file was recorded at 5 MHz sampling rate, with INT8 samples in I-Q interleaved format and had nominal GPS signal conditions. The timing scale used was in nanoseconds. Also, conventional tracking integration lengths of 1 millisecond were used. Table 5 shows offline results.

**Table 5. Offline receiver tracking performance for different acceleration factors**

| SDR Tracking Time (nsec) | Avg. time per epoch per channel (nsec) | No. of real-time tracking channels* |
|---|---|---|
| Base version | 75,920.97 | 13.17 |
| Multiple-producer/consumer loop feature | 75,787.05 | 13.19 |
| Parallelizable loops feature | 40,258.45 | 24.84 |
| SIMD feature | 11,125.70 | 89.88 |

It can be seen that the feature which increases acceleration the most is the SIMD feature version, which adds all LabVIEW features together. The acceleration factor can be seen up to 6.8 times faster than the base version. Also, up to 89 tracking channels could be theoretically run continuously, although the receiver can only run 12 tracking channels. This translates into added overhead for experimenting other algorithms for research purposes.

For online mode, other metrics were measured to see receiver performance, such as CPU load, memory usage, number of threads, number of channels in real-time mode and max number of channels in any sampling rate. Table 6 shows online results. These results were obtained by running receiver with simulator to output always 12 SVs visible. Also, the numbers are averaged after running the receiver for 50 minutes on each different version.

**Table 6. Online receiver tracking performance for different acceleration factors**

| Online testing at 5 MHz | CPU load (%) | Memory (KB) | No. of threads | No. of channels in real-time @ 5 MHz | Max. no. of channels in real-time @ Fs |
|---|---|---|---|---|---|
| Base version | 15.99 | 167,864.88 | 34 | 11 | 11 @ 5 MHz |
| Multiple-prod./cons. loop feature | 15.52 | 155,346.13 | 39 | 11 | 12 @ 4 MHz |
| Parallelizable loops feature | 22.73 | 155,248.85 | 39 | 12 | 9 @ 10 MHz |
| SIMD feature | 9.85 | 155,613.47 | 39 | 12 | 8 @ 25 MHz |

Each version easily achieved near 12 channels which is the hard-coded limit of tracking channels on the receiver. For experimental purposes, other sampling frequencies were tested to stretch receiver capabilities to its maximum. It can be seen that the SIMD feature version was able to operate at 25 MHz and tracking 8 real-time channels without performance degradation. A reason some values on this table, such as number of tracking channels at 5 MHz, could be lower is because real-time operation entitles other critical resources such as front-end data fetching which could translate into extra receiver computational overhead as expected.

## 5 CONCLUSION

This paper presented a new generation of LabVIEW-based real-time GPS receivers which relied on DLL integration of C/C++ baseband functions. The hardware testbed was presented as an option for compactness and mobility. Several LabVIEW-based features were explored as well as other acceleration factors that achieve real-time operation in the receiver. Some aspects such as advanced acquisition algorithms, SIMD technologies, optimized libraries, and other factors were discussed. LabVIEW-based features such as DLL integration, parallelizable loops, and built-in visualizations were shown as leverage for fast prototyping of a complete GPS receiver operating in real-time. It was shown that LabVIEW provides inherent multithreading schemes such as the producer/consumer loop architecture, and also deals with data flow handling. LabVIEW also provides a built-in interface between the front-end and the software receiver. Performance numbers showed that with all features, receiver experienced up to 6.8 times faster than without proposed features. Also, as much as 89 tracking channels were able to run theoretically in offline mode, translating into future computational overhead space in the receiver for algorithm testing and other research purposes. In online mode, the receiver achieved 8 real-time tracking channels at 25 MHz sampling rate with all proposed acceleration factors, thus demonstrating a competitive LabVIEW-based receiver.